\documentclass[prl,twocolumn,superscriptaddress,showpacs]{revtex4}
\usepackage{amsmath}
\usepackage{graphicx}
\usepackage{bm}
\usepackage[T1]{fontenc}
\usepackage[latin1]{inputenc}
\usepackage{times}
\usepackage{amssymb}

\begin{document}

\title{Measuring the Transmission Matrix in Optics :\\ An Approach to the Study and Control of Light Propagation in Disordered Media}

\author{S. M. Popoff, G. Lerosey, R. Carminati, M. Fink, A.C. Boccara, S. Gigan}
\affiliation{Institut Langevin, ESPCI ParisTech, CNRS UMR 7587, ESPCI, 10 rue Vauquelin, 75005 Paris, France.}

\begin{abstract}
We introduce a method to experimentally measure the monochromatic transmission matrix of a complex medium in optics. This method is based on a spatial phase modulator together with a full-field interferometric measurement on a camera. We determine the transmission matrix of a thick random scattering sample. We show that this matrix exhibits statistical properties in good agreement with random matrix theory and allows light focusing and imaging through the random medium. This method might give important insights into the mesoscopic properties of complex medium.
\end{abstract}

	
\maketitle

The propagation of waves in heterogeneous media, especially in the multiple scattering regime, is a very fundamental problem of physics with numerous applications ranging from solid-state physics and optics,  to acoustics and electromagnetism \cite{sheng95}. 
The behavior or such media is usually described by their average macroscopic properties such as the scattering and absorption mean free paths, and via spectral or temporal properties, such as spectral resonances or confinement times. Some of those quantities can be obtained through intensity measurements using for example speckle pattern correlations \cite{li1994correlation} or coherent back scattering experiments \cite{wolf1985weak,PhysRevLett.55.2692}.

A deeper approach for the study of complex media lies in the transmission matrix (TM) retrieval. This matrix is a subpart of the usual scattering matrix as defined in \cite{RevModPhys.69.731} for instance. Within this framework, the Green's function between an array of sources and an array of sensors is recorded in transmission. The knowledge of the TM brings more fundamental insight on the medium. One can for instance extract from the TM the single and the multiple scattering component \cite{aubry2009random}, the backscattering cone, and the field-field correlations \cite{mello1988macroscopic,sebbah2002spatial}. The distribution of the singular values of the TM should also be related to diffusion properties and might exhibit coherence effect beyond the diffusive transport regime  such as weak and strong localization effects.

Furthermore, from an operative point of view, the knowledge of the TM of a complex medium offers new and exciting possibilities. For instance, using Time Reversal (TR), which is a broadband version of Phase Conjugation (PC), it has been shown that multiple scattering, far from limiting wave manipulation through a random medium, can in fact greatly enhance it. Such approaches, which are based on the reciprocity and the reversibility of the wave equations \cite{carminati2000reciprocity}, have proved very useful in various areas ranging from focusing to imaging or telecommunication, in acoustic \cite{FinkPhysTo97,PhysRevLett.75.4206,PhysRevLett.90.014301}, electromagnetic \cite{lerosey2007focusing,PhysRevLett.92.193904} or even seismology \cite{shapiro2005high}.

For acoustic (resp. electromagnetic) waves, transducers (resp. antennas) are natural local receivers and emitters in phase and amplitude, and oscillation frequencies are compatible with electronic detection. Therefore, the TM can be straightforwardly measured. In contrast, reconstructing the transmission matrix of a complex medium is still an elusive problem in the optical domain.

Nonetheless some recent experiments have demonstrated that it is possible to manipulate light at a mesoscopic level in a complex medium in order to focus light through \cite{vellekoop2007focusing} or in \cite{vellekoop2008demixing} a scattering medium, as well as couple efficiently to the open channels of a disordered sample \cite{vellekoop2008universal}. These experiments were made possible by the emergence of Spatial Light Modulators, which allow to control the phase of an optical field over a million pixels, and were based on algorithmic procedures which only give a limited information on the physical behavior of the medium.

In the present paper, using a spatial phase modulator alongside a full-field interferometric method, we demonstrate that we are able to measure the monochromatic transmission matrix of a random multiple scattering medium. We first detail the experimental setup of the TM evaluation method. In a second part, we start by showing that the knowledge of the TM allows one to reproduce the results of \cite{vellekoop2007focusing}, and make the link with TR experiments \cite{PhysRevLett.92.193904,PhysRevLett.75.4206}. Then we demonstrate the potential of our approach with the detection of an unknown object after propagation through the scattering media. Finally, we study the statistical properties of the singular values of the TM, and show that it follows a quarter circle law typical of a random wavefield.

We define the mesoscopic TM of an optical system for a given wavelength as the matrix $K$ of the complex coefficients $k_{mn}$ connecting the optical field (in amplitude and phase) in the $m^{th}$ output free mode to the one in the $n^{th}$ input free mode. Thus, the projection $E^{out}_m$ of the outgoing optic field on the $m^{th}$ free mode is given by 
\begin{equation}
E^{out}_m = \sum_n{k_{mn}E^{in}_n}
\label{eqE}
\end{equation}

Where $E^{in}_n$ is the complex amplitude of the optical field in the $n^{th}$ incoming free mode.

The sample under study is an opaque 80 $\pm$ 25 $\mu m$ thick deposit of ZnO (Sigma-Aldrich 96479) on a standard microscope glass slide. The experimental setup consists of a diode-pumped solid-state single longitudinal mode laser source at 532 nm (Laser Quantum Torus), a modulation part to illuminate the sample with a controlled wave front and a detection part to measure the optical intensity transmitted by the scattering medium. The laser beam is first expanded then spatially modulated by a liquid cristal Spatial Light Modulator (SLM). This modulator is a twisted nematic liquid cristal on silicon device (Holoeye LC-R 2500). Choosing a suitable combination of incident and analysed polarizations, we achieve an almost phase-only modulation~\cite{davis2002phasor} of the collimated incident beam with a 2$\pi$ modulation in phase and a maximum residual intensity modulation below 10\%. The surface of the SLM is imaged on the pupil of a 20x objective with a numerical aperture (NA) of 0.5, thus a pixel of the SLM matches a wave vector at the entrance of the scattering medium. The beam is focused at one side of the sample and the output intensity speckle is imaged 0.3 mm from the surface of the sample by a 40x objective (NA = 0.85) onto a 10-bit CCD camera (AVT Dolphin F-145B). The speckle is stationnary well over the measurement time (several minutes).

\begin{figure}[ht]
\center
\includegraphics[width=0.45\textwidth]{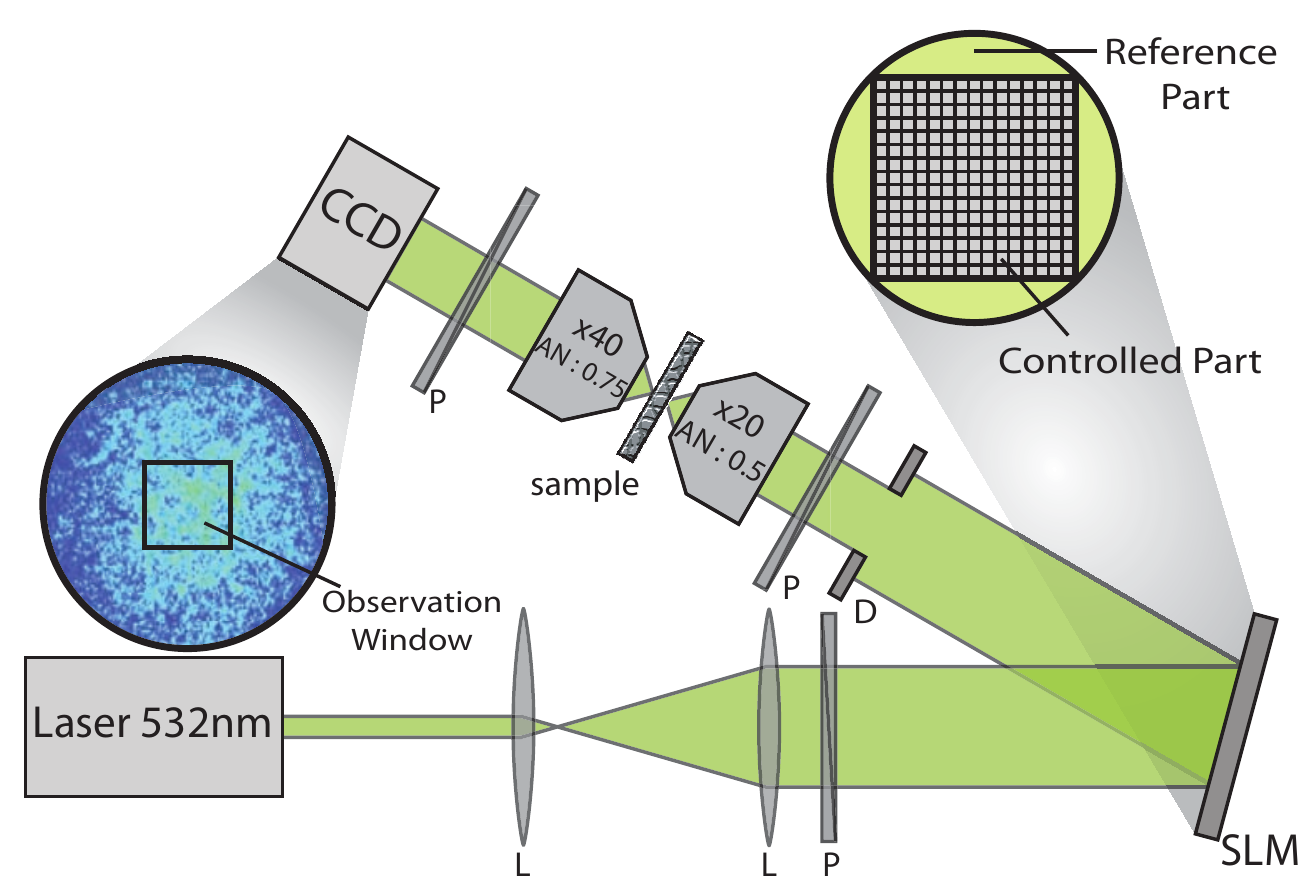}
\caption{ Schematic of the apparatus. The 532 nm laser is expanded and reflected off a SLM. Polarization optics select a phase only modulation mode. The modulated beam is focused on the multiple-scattering sample and the output intensity speckle pattern is imaged by a CCD-camera. L, lens. P, polarizer. D, diaphragm.}
\label{Setup}
\end{figure}

We choose the size of the input and output independent pixels to have a perfect matching in size between a pixel and a mode, in particular a CCD pixel has the size of a speckle grain. Thus, in our setup, the input and output modes are the SLM and the CCD pixels respectively, and the TM corresponds to the system comprised of both the scattering sample and the optical system between the SLM and the CCD Camera. From now on, we fixed both the number of controlled segments on the SLM and the number of subdivisions measured on the observation windows of the CCD to $N=256$.

Measuring a TM in optics raises several difficulties coming from the impossibility of having direct access to the amplitude and phase of the optical field. Generally, one can have access to the complex optic field using interferences with a known wavefront and a full field "four phases method" \cite{dubois2002high}. For any input vector, if the global phase is shifted by a constant $\alpha$, the intensity in the $m^{th}$ output mode is given by\begin{eqnarray}
&I_m^\alpha &= |E^{out}_m|^2 = |s_m + \sum_n{e^{i \alpha}k_{mn}E^{in}_n}|^2 \\ \nonumber
  &= &|s_m|^2+|\sum_n{e^{i\alpha}k_{mn}E^{in}_n|^2}
 +2\Re{\left(e^{i\alpha}\overline{s}_m \sum_n{k_{mn}E^{in}_n }\right)} \nonumber
\label{equI}
\end{eqnarray}

where $s_m$ is the complex amplitude of the optical field used as reference in the $m^{th}$ output mode.

Thus, if we inject the $n^{th}$ input mode and we measure $I^{0}_m$,$I^{\frac{\pi}{2}}_m$,$I^{\pi}_m$ and $I^{\frac{3\pi}{2}}_m$, respectively the intensities in the $m^{th}$ outgoing mode for $\alpha=0$, $\pi/2$, $\pi$ and $3\pi/2$ gives :	
\begin{eqnarray}
\frac{\left(I^{0}_m-I^{\pi}_m\right)}{4}+ i \frac{\left(I^{\frac{3\pi}{2}}_m-I^{\frac{\pi}{2}}_m\right)}{4} = \bar{s}_m k_{mn}
\end{eqnarray}

For practical reasons, we choose the Hadamard basis as input basis over the canonical one. In the Hadamard basis, all elements are either +1 or -1 in ampltiude, which perfectly fits with the use of a phase-only SLM. Another advantage is that it also maximizes the intensity of the received wavefield and consequently improves the experimental signal to noise ratio (SNR). For all Hadamard basis vectors, the intensity is measured on the canonical basis of the pixels on the CCD camera and an observed transmission matrix $K_{obs}$ is acquired, which is related to the real one $K$ by
\begin{equation}
K_{obs} = K\times S_{ref}
\label{kobs}
\end{equation}

where $S_{ref}$ is a diagonal matrix representing the whole static reference wavefront in amplitude and phase. Ideally, the reference wavefront should be a plane wave to directly have access to the $K$ matrix. In this case, all $s_{m}$ are constant and $K_{obs}$ is directly proportional to $K$. However this requires the addition of a reference arm to the setup, and requires interferometric stability. To have the simplest experimental setup and a higher stability, we modulate only 65 \% of the wavefront going into the scattering sample (this  correponds to the square inside the pupil of the microscope objective as seen in Figure \ref{Setup}), the speckle coming from the 35 \% static part being our reference. $S_{ref}$ is now unknown and no longer constant along its diagonal. Nevertheless, since $S_{ref}$ is stationnary over time, we can measure the response of all input vectors on the $m^{th}$ output pixel as long as the reference speckle is bright enough on the considered modes. We will quantify the effect of the reference speckle and show that neither does it impair our ability to focus or image using the TM, nor does it affect the statistical properties of the TM.

A good way to confirm the physical relevance of the TM is to use it to focus light on any desired outgoing mode. In \cite{PhysRevLett.92.193904,PhysRevLett.75.4206}, it was demonstrated that using time reversal, one can take advantage of multiple scattering to focus on tight spots, thanks to the reversibility of the wave equation. In this configuration, the Green's functions between the array of sources and a point at the output of the disordered medium is sent reversed in time, and the waves generated converge naturally towards the targeted spot. TR being a matched filter \cite{PhysRevLett.75.4206}, the energy is maximized both temporally and spatially at the intended location. The monochromatic equivalent of TR is phase conjugation, which can be straightfully done using our acquired TM. We expect similar focusing results as the ones of \citet{vellekoop2007focusing}, which were obtained through a procedure that ensures an optimum phase modulated wavefront maximizing the energy of a given mode. Denoting $E^{target}$ the output target vector, the input vector that approximates the desired pattern at the output for a perfect phase conjugated focusing is given by
\begin{equation}
E^{in} =  K^{\dag}E^{target}
\end{equation}

where $^\dag$ stands for the conjugate transpose. Thus, the theoretical effective output vector is $E^{eff} =  O^{foc}E^{target}$ where $O^{foc}=KK^{\dag}$ is the so called time reversal operator \cite{prada1994time}. The result of the theoretical focusing on the $m^{th}$ outgoing mode is the $m^{th}$ column of $O^{foc}$. Since our setup is limited to phase modulation only, and given the fact that we do not acquire $K$ but $K_{obs}$, the input vector for a given output target reads 
\begin{equation}
E^{in} = \frac{ K_{obs}^{\dag}E^{target}}{| K_{obs}^{\dag}E^{target}|}
\label{pseudoconj}
\end{equation}

\begin{figure}[ht]
\center
\includegraphics[width=0.45\textwidth]{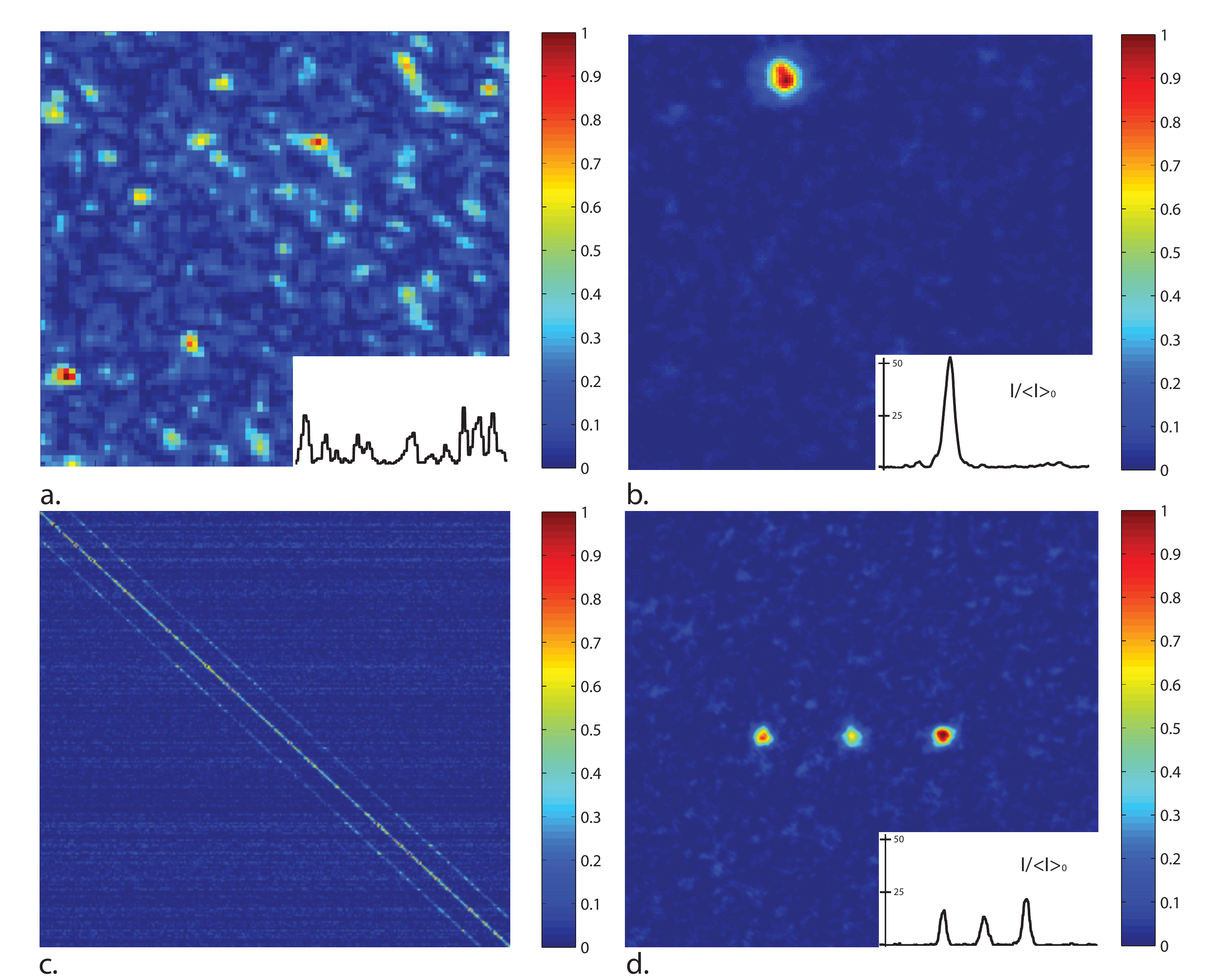}
\caption{Experimental results of focusing. \textbf{a.} initial aspect of the output speckle. \textbf{b.} we measure the TM for 256 controlled segments and use it to perform phase conjugation. \textbf{c.} norm of the focusing operator $O_{norm}^{foc}$. \textbf{d.} example of focusing on several points. (the insets show intensity profiles along one direction)}
\label{Foc}
\end{figure}

We use the setup shown in Figure \ref{Setup} to record the transmission matrix of our system, which is done in approximately 3 minutes (and require $4N$ measurements), comparable to the time needed to perform an iterative focusing as described in \cite{vellekoop2007focusing}. A speckle spot is chosen as the intended target, and we use formula \ref{pseudoconj} to perform phase conjugation. The results obtained are shown in Figure \ref{Foc}. 

In the single spot focusing experiment (Figure \ref{Foc}.b), we estimate the experimental SNR defined by the ratio of the intensity at the focus to the mean intensity of the speckle outside. The result of the focusing on this target shows a $\text{SNR}_{exp} = 54$ which \textit{in par} the results of the focusing technique used by Vellekoop \textit{et al.} \cite{vellekoop2007focusing}. 

It is important to quantify the effect of both the phase-only modulation and the reference speckle used to acquire the TM of our complex medium. Indeed, rather then $O^{foc}$, the effective focusing operator includes a normalization to take into account the input phase-only modulation. For a single spot focusing, $O_{norm}^{foc}=K_{obs}K_{norm}^{\dag}$ where $k^{norm}_{ij}=k_{ij}/|k_{ij}|$. In the representation of the focusing operator $O_{norm}^{foc}$ (Figure \ref{Foc}.c), each line corresponds to the expected output intensity image resulting while focusing on any point of the CCD's observation window. The strong diagonal proves our ability to focus light on any of these spots. This is examplified in Figure \ref{Foc}.d, which shows focusing on 3 spots. It is worth noting that $O_{norm}^{foc}$ presents weaker sub-diagonals which are due to correlations between neighbouring pixels.
Finally we would like to emphasize that the strong advantage of this focusing technique, compared to a sequential algorithm, is that it is not necessary to acquire anymore information to change the focusing spot or to focus on multiple spots in one step.

From a theoretical point of view, monochromatic phase conjugation focusing through a multiple scattering medium has been studied by \citet{derode2001scattering}. The general formulation for the efficiency of this focusing is SNR $\approx N_{grains}$ the total number of "information grains", or degrees of freedom. In our case, this number is the number of independent pixels (or channels) which is the number $N$ of pixels controlled. In our experiment, the use of a reference speckle limits the number of degrees of freedom that are controlled. Denoting $\gamma$ the fraction of the SLM used to acquire the TM, and using the theoretical formalism developed in \cite{derode2001scattering} the effective SNR including the effect of the reference writes
\begin{equation}
\text{SNR}_{ref} = \pi/4 \left(\gamma N + {(1-\gamma)}/{\gamma}\right) \approx 0.8 \gamma N   \quad   \forall N \gg 1
\end{equation}

Where $(1-\gamma)/\gamma$ is the effect of the reference speckle which contributes as noise at the focal spot, and the prefactor is due to the phase-only modulation \cite{vellekoop2007focusing}, in agreement with a similar experiment in acoustics \cite{derode1999ultrasonic}. The experimental SNR of the focusing shown in Figure \ref{Foc} \textbf{b.} is 41\% of SNR$_{ref}$, with $\gamma = 65\%$.

We have shown that the TM of the medium is a powerful tool that can be used to focus through a medium, but we want to emphasize here that that it extends well beyond focusing. An example is the detection of an amplitude or phase object placed at the entrance of the scattering sample. Due to the spatial reciprocity of wave propagation inside the scattering medium, the detection issue is reciprocal to the focusing one. Once the matrix is measured, one can retrieve the initial input $E_{obj}$ signal from the output speckle $E_{out} = K_{obs}.E_{obj}$ with the focusing operator $O^{foc}$. Indeed, the reconstructed image reads
\begin{equation}
E_{img} \approx K_{obs}^{\dag}E_{out} = O^{foc\dag} E_{obj} =O^{foc}E_{obj}
\end{equation}

\begin{figure}[ht]
\center
\includegraphics[width=0.4\textwidth]{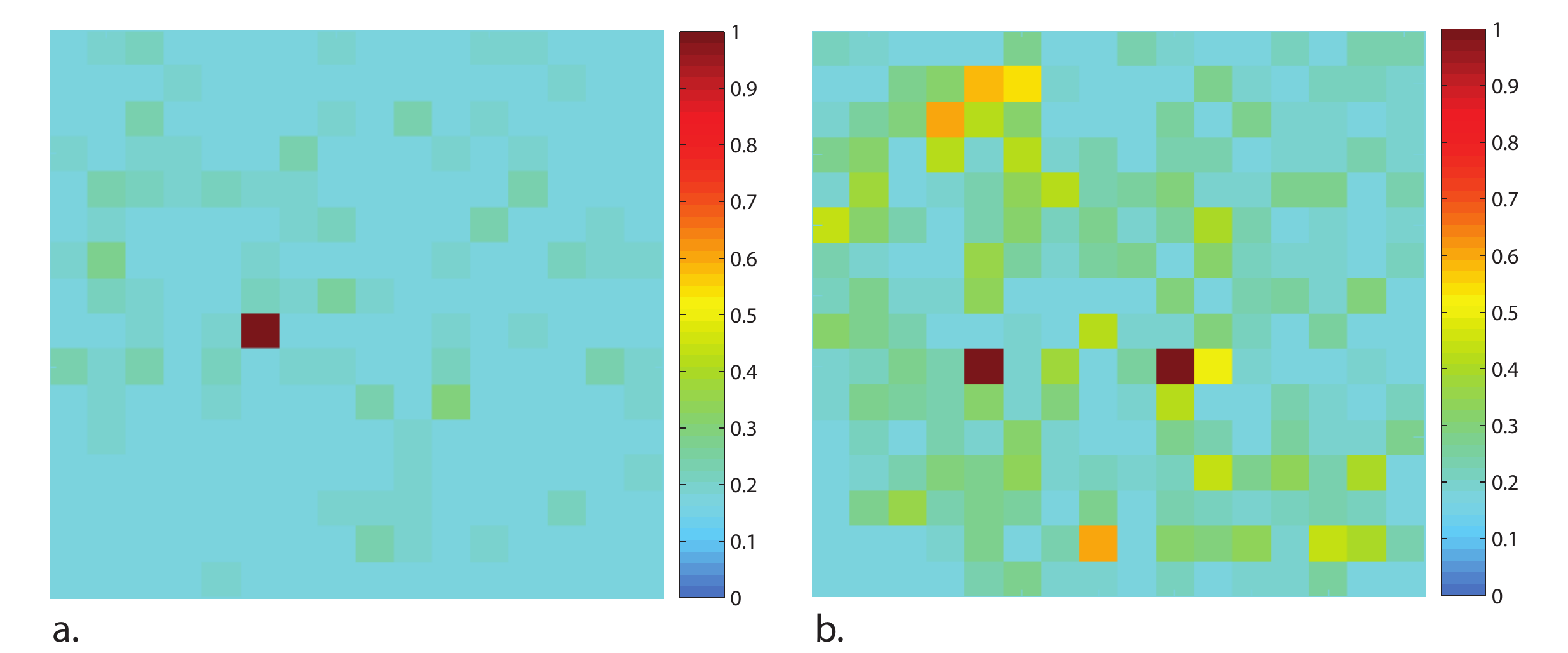}
\caption{Experimental results of detection of an object placed before the scattering sample. Image \textbf{a.} (resp \textbf{b.}) shows the result of one spot (two spots) detection for 256 controlled segments on the SLM.}
\label{Detection}
\end{figure}

Which is almost equivalent to the result obtained in a focusing scheme since $O^{foc}$ as the same statistical properties than $O_{norm}^{foc}$ but without the phase-only effects. In order to demonstrate the validity of this approach we performed the following detection experiment: instead of placing an object between the SLM and the medium, the object is generated by the SLM itself using two phase masks $E_{obj}^{(1)}$ and $E_{obj}^{(2)}$. The mask $E_{obj}^{(1)}$ is a plane phase and $E_{obj}^{(2)}$ is obtained  by flipping the phase of $N_{obj}$ pixels from 0 to $\pi$. The phase and amplitude $E^{(1)}_{out}$ and $E^{(2)}_{out}$ of the output speckles are measured. The virtual object $E_{obj}^{(1)}-E_{obj}^{(2)} $  is an amplitude object, recovered by calculating $E_{img} = K_{obs}(E_{out}^{(1)}-E_{out}^{(2)} )$. It gives the images that would be obtained if an opaque screen drilled with $N_{obj}$ holes of a speckle grain size was placed at the pupil of the input microscope objective. The results of this operation are shown in Figure \ref{Detection} for $N_{obj}=1$ and $N_{obj}=2$. It is clear that a simple object placed at the input of the sample can be reconstructed after propagation, regardless of the randomness of the medium. Considering that this imaging method is reciprocal of the focusing one, the detection is achieved with the same SNR.

These experimental focusing and detection experiments through a multiple scattering sample are clear demonstration that the measured transmission matrix is physical, i.e  effectively links the input optic field to the output one, both in amplitude and phase. Theoretically, this matrix is related to the scattering matrix and should be an extremely powerful tool for the study of random and complex media. As a first step in this direction, we performed the simplest analysis of the TM: a singular value decomposition. Indeed, provided that there is no ballistic wave in the field at the output of the sample, Random Matrix Theory (RMT) predicts that for those matrices the statistical distribution $\rho(\widetilde{\lambda})$ follows the so-called "quarter circle law" where $\widetilde{\lambda}$ are the singular values (SV) normalized by the total intensity \cite{marcenko1967distribution}. 

\begin{figure}[ht]
\center
\includegraphics[width=0.35\textwidth]{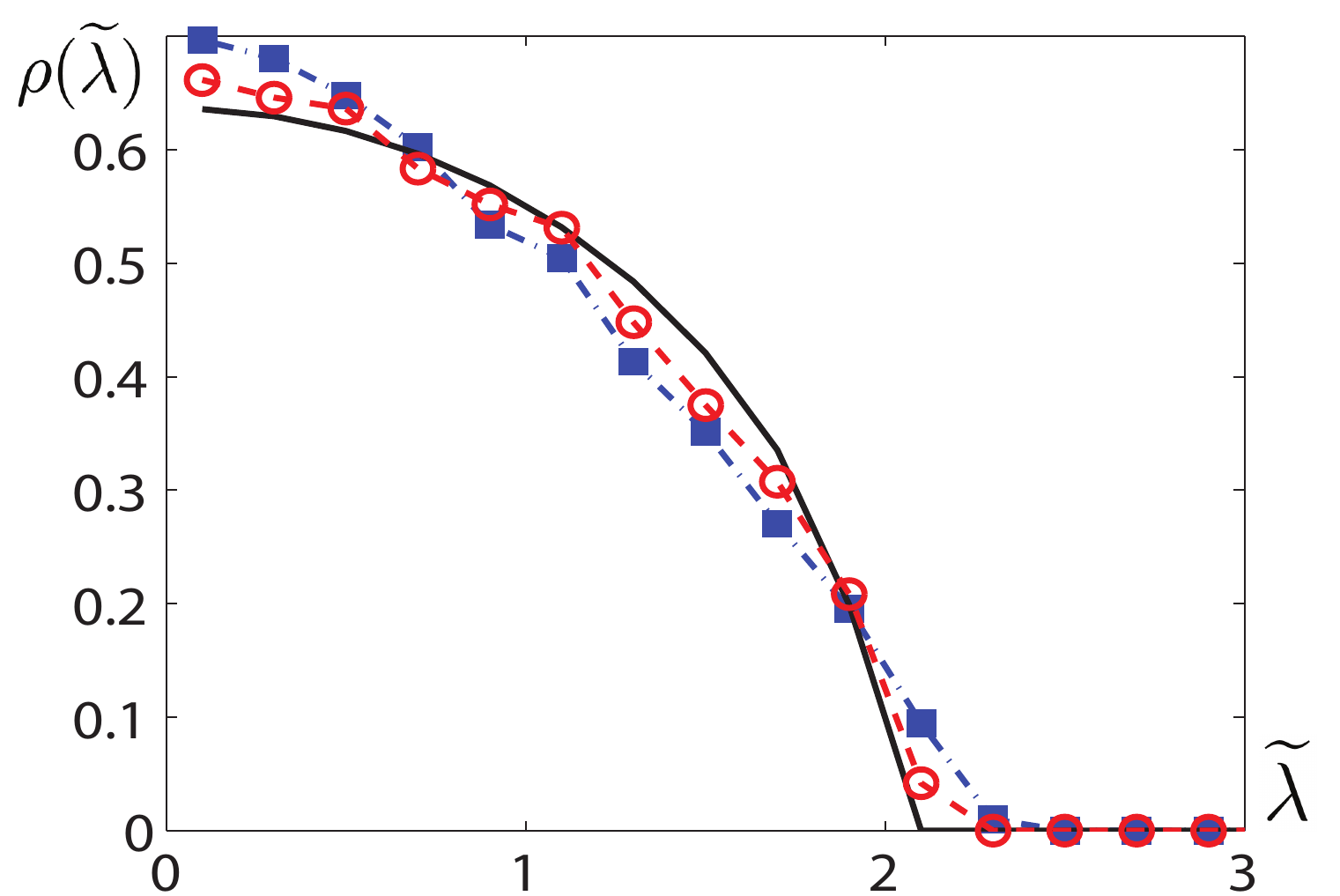}
\caption{Singular values distribution of the experimental transmission matrices obtained by averaging over 16 realizations of disorder. The solid line is the  quarter-circle law predicted for random matrices.  With $\blacksquare$ the matrix filtered to remove the reference amplitude contribution and with $\circ$ the matrix obtained by filtering and removing neighboring elements to eliminate inter-element correlations. }
\label{Vp}
\end{figure}

The effect of the diagonal matrix $S_{ref}$ on $K$ is to multiply each line by a value $s_{jj}$. This changes the variance of the TM which affects the SV distribution. To remove this effect, we estimate the absolute values of $S_{ref}$ matrix by averaging the $K_{obs}$ matrix along every lines. We obtain the diagonal matrix $S_{abs}$ defined by $s^{abs}_{jj} = \langle|k^{obs}_{ij}|\rangle_i \propto |s_{jj}|$ for $\langle |k_{ij}| \rangle = 1$. $S_{abs}$ has has no zero on the diagonal thanks to the finite pixel size. We filter the $K_{obs}$ matrix and obtain :
\begin{equation}
K_{fil} = K\times \frac{S_{ref}}{S_{abs}} \propto K\times \Sigma_{\phi}
\end{equation}

Now, $\Sigma_{\phi}$ is a unitary matrix since it is a diagonal matrix of complex number of norm $1$, corresponding to the unknown phase of the reference speckle. Thus the singular values of $K_{fil}$ are the same as those for $K$. We finally remove one element out of two to remove inter-element correlations between nearby pixels. Results of SV distribution on the observed matrix after filtering and under-sampling are shown in Figure \ref{Vp}, the final matrix obtained follows correctly the expected quarter-circle law. This distribution differs from the bimodal one predicted by \cite{pendry1999} since there is no total energy conservation conditions as the transmission matrix measured is a small part of the whole scattering matrix

To conclude, we have demonstrated how the optical transmission matrix of a multiple scattering media can be measured both in phase and amplitude using a spatial light modulator and a full-field interferometric measurement. We proved the validity of the measured TM through focusing and detection experiments. Finally, the singular values distribution was shown to follow a quarter circle law, characteristic of random matrices. We believe that this technique will have many applications ranging from the control of wavefield on detection and imaging in random and scattering media to very fundamental problems associated with the propagation of wave in disordered systems.

We wish to thank Alexandre Aubry, Samuel Grésillon et Aurélien Peilloux for inspiring discussions. This work was made possible by financial support from "Direction Générale de l'Armement" (DGA) and BQR funding from Université Pierre et Marie Curie and ESPCI.

\end{document}